\documentclass[preprint2]{aastex}
\usepackage{epstopdf}

\shorttitle{WISE Interactive Session Results}
\shortauthors{Faherty et al.}

\begin{document}

\title{Results from the Wide-field Infrared Survey Explorer (WISE) Future Uses Session at the WISE at 5 Meeting}
\author{Jacqueline K. Faherty\altaffilmark{1,2,3}, K. Alatalo\altaffilmark{4},
L. D. Anderson\altaffilmark{5},
Roberto J. Assef\altaffilmark{6},
Daniella C. Bardalez Gagliuffi\altaffilmark{7},
Megan Barry\altaffilmark{8},
Dominic J. Benford\altaffilmark{9},
Maciej Bilicki\altaffilmark{10,11},
Ben Burningham\altaffilmark{12,13},
Damian J. Christian\altaffilmark{8},
Michael C. Cushing\altaffilmark{14},
Peter R. Eisenhardt\altaffilmark{15},
Martin Elvisx\altaffilmark{29},
S. B. Fajardo-Acosta\altaffilmark{4},
Douglas P. Finkbeiner\altaffilmark{29},
William J. Fischer\altaffilmark{9},
William J. Forrest\altaffilmark{17},
John Fowler\altaffilmark{4},
Jonathan P. Gardner\altaffilmark{18},
Christopher R. Gelino\altaffilmark{4},
V Gorjian\altaffilmark{15},
Carl J. Grillmair\altaffilmark{19},
Mariusz Gromadzki\altaffilmark{20},
Kendall P. Hall\altaffilmark{21},
\v{Z}eljko Ivezi\'{c}\altaffilmark{22},
Natsuko Izumi\altaffilmark{23},
J. Davy Kirkpatrick\altaffilmark{4},
Andr\'as Kov\'acs\altaffilmark{24,25},
Dustin Lang\altaffilmark{26,27},
David Leisawitz\altaffilmark{16},
Fengchuan Liu\altaffilmark{15},
A. Mainzer\altaffilmark{15},
Katarzyna Malek\altaffilmark{28},
G\'abor Marton\altaffilmark{30},
Frank J. Masci\altaffilmark{4},
Ian S. McLean\altaffilmark{31},
Aaron Meisner\altaffilmark{29},
Robert  Nikutta\altaffilmark{32},
Deborah L. Padgett\altaffilmark{19},
Rahul Patel\altaffilmark{33},
L. M. Rebull\altaffilmark{34},
J. A. Rich\altaffilmark{4,35},
Frederick A. Ringwald\altaffilmark{21},
Marvin Rose\altaffilmark{29},
Adam C. Schneider\altaffilmark{14},
Keivan G Stassun\altaffilmark{36, 37},
Daniel Stern\altaffilmark{15},
Chao-Wei Tsai\altaffilmark{15},
Feige Wang\altaffilmark{38, 39},
Madalyn E. Weston\altaffilmark{40},
Edward L. (Ned)  Wright\altaffilmark{31, 41},
Jingwen Wu\altaffilmark{13},
Jinyi Yang\altaffilmark{38, 39}}

\altaffiltext{         1	}       {Department of Terrestrial Magnetism, Carnegie Institution of Washington, Washington, DC 20015, USA; jfaherty@carnegiescience.edu }
\altaffiltext{         2	}       {Department of Astrophysics,  American Museum of Natural History, Central Park West at 79th Street, New York, NY 10034}
\altaffiltext{         3	}       {Hubble Fellow}
\altaffiltext{		4	}	{Infrared Processing \& Analysis Center, California Institute of Technology, Pasadena, CA 91125, USA	}
\altaffiltext{		5	}	{Department of Physics and Astronomy, West Virginia University, Morgantown, WV 26506	}
\altaffiltext{		6	}	{"N\'ucleo de Astronom\'ia de la Facultad de Ingenier\'ia, Universidad Diego Portales, Av. Ej\'ercito Libertador 441, Santiago, Chile.}
\altaffiltext{		7	}	{Center for Astrophysics and Space Sciences, University of California, San Diego, 9500 Gilman Dr., Mail Code 0424, La Jolla, CA 92093, USA.	}
\altaffiltext{		8	}	{Department of Physics and Astronomy, California State University Northridge, 18111 Nordhoff Street, Northridge, CA 91330, USA	}
\altaffiltext{		9	}	{NASA Goddard Space Flight Center, Code 665, Greenbelt, MD 20771, USA	}
\altaffiltext{		10	}	{Department of Astronomy, University of Cape Town, Private Bag X3, 7701 Rondebosch, South Africa	}
\altaffiltext{		11	}	{Kepler Institute of Astronomy, University of Zielona Gora, ul. Szafrana 2, 65-516 Zielona Gora, Poland	}
\altaffiltext{		12	}	{Centre for Astrophysics Research, Science and Technology Research Institute, University of Hertfordshire, Hatfield AL10 9AB, UK	}
\altaffiltext{		13	}	{NASA Ames Research Center, Mail Stop 245-3, Moffett Field, CA 94035, USA	}
\altaffiltext{		14	}	{Department of Physics and Astronomy, The University of Toledo, 2801 West Bancroft Street, Toledo, OH 43606, USA	}
\altaffiltext{		15	}	{Jet Propulsion Laboratory, California Institute of Technology, 4800 Oak Grove Drive, Pasadena, CA 91109, USA	}
\altaffiltext{		16	}	{NASA Goddard Space Flight Center, Code 605, Greenbelt, MD 20771, USA	}
\altaffiltext{		17	}	{Department of Physics and Astronomy, University of Rochester, Rochester, NY 14627-0171	}
\altaffiltext{		18	}	{Laboratory for Observational Cosmology, Astrophysics Science Division, Code 665, Goddard Space Flight Center, Greenbelt, MD 20771, USA	}
\altaffiltext{		19	}	{Spitzer Science Center 1200 E. California Blvd., Pasadena, CA 91125	}
\altaffiltext{		20	}	{Millennium Institute of Astrophysics, Santiago, Chile Istitiuto de F\'{i}sica y Astronom\'{i}a, Universidad de Valpara\'{i}so, ave. Gran Breta\~{n}na, 1111, Casilla 5030, Valpara\'{i}so, Chile	}
\altaffiltext{		21	}	{Department of Physics California State University, Fresno 2345 E. San Ramon Ave., M/S MH37 Fresno, CA 93710 U.S.A.	}
\altaffiltext{		22	}	{University of Washington Department of Astronomy Box 351580 Seattle, WA 98195	}
\altaffiltext{		23	}	{Institute of Astronomy, School of Science, The University of Tokyo	}
\altaffiltext{		24	}	{Institut de F\'isica d'Altes Energies, Universitat Aut\'onoma de Barcelona, E-08193 Bellaterra (Barcelona), Spain	}
\altaffiltext{		25	}	{MTA-ELTE EIRSA "Lend\"ulet" Astrophysics Research Group, 1117 P\'azm\'any P\'eter s\'et\'any 1/A Budapest, Hungary	}
\altaffiltext{		26	}	{Department of Physics \& Astronomy, University of Waterloo, 200 University Avenue West, Waterloo, Ontario, Canada N2L 3G1	}
\altaffiltext{		27	}	{McWilliams Center for Cosmology, Department of Physics, Carnegie Mellon University, 5000 Forbes Ave, Pittsburgh, PA 15213	}
\altaffiltext{		28	}	{National Centre for Nuclear Research, ul. Hoza 69, 00-681 Warszawa, Poland	}
\altaffiltext{		29	}	{Harvard Smithsonian Center for Astrophysics, 60 Garden St., Cambridge, MA 02138, USA	}
\altaffiltext{		30	}	{Konkoly Observatory Research Centre for Astronomy and Earth Sciences Hungarian Academy of Sciences 1121 Budapest Konkoly Thege Mikl\'os \'ut 15-17 Hungary	}
\altaffiltext{		31	}	{Department of Physics and Astronomy, UCLA, Los Angeles, CA 90095-1562, USA	}
\altaffiltext{		32	}	{Instituto de Astrof\'isica, Facultad de F\'isica, Pontificia Universidad Cat\'olica de Chile, 306, Santiago 22, Chile	}
\altaffiltext{		33	}	{Department of Physics and Astronomy State University of New York at Stony Brook	}
\altaffiltext{		34	}	{Infrared Science Archive (IRSA) and Spitzer Science Center (SSC), Infrared Processing and Analysis Center (IPAC), 1200 E. California Blvd, MS 314-6, Pasadena, CA, 91125.	}
\altaffiltext{		35	}	{Observatories of the Carnegie Institution of Washington, 813 Santa Barbara Street, Pasadena, CA 91101, USA	}
\altaffiltext{		36	}	{Fisk University, Department of Physics, 1000 17th Ave. N., Nashville, TN 37208, USA	}
\altaffiltext{		37	}	{Vanderbilt University, Department of Physics \& Astronomy, VU Station B 1807, Nashville, TN 37235, USA }
\altaffiltext{		38	}	{Department of Astronomy, School of Physics, Peking University, Beijing 100871, China	}
\altaffiltext{		39	}	{Steward Observatory, University of Arizona, 933 North Cherry Avenue, Tucson, AZ 85721, USA	}
\altaffiltext{		40	}	{Department of Physics \& Astronomy, University of Missouri - Kansas City, Kansas City, MO 64110, USA	}
\altaffiltext{		41	}	{David Saxon Presidential Chair in Physics  	}

\begin{abstract}
During the ``WISE at 5: Legacy and Prospects'' conference in Pasadena, CA -- which ran from February 10 - 12, 2015 -- attendees were invited to engage in an interactive session exploring the future uses of the Wide-field Infrared Survey Explorer (WISE) data. The 65 participants -- many of whom are extensive users of the data -- brainstormed the top questions still to be answered by the mission, as well as the complementary current and future datasets and additional processing of WISE/NEOWISE data that would aid in addressing these most important scientific questions. The results were mainly bifurcated between topics related to extragalactic studies (e.g. AGN, QSOs) and substellar mass objects. In summary, participants found that complementing WISE/NEOWISE data with cross-correlated multiwavelength surveys (e.g. SDSS, Pan-STARRS, LSST, Gaia, Euclid, etc.) would be highly beneficial for all future mission goals. Moreover, developing or implementing machine-learning tools to comb through and understand cross-correlated data was often mentioned for future uses. Finally, attendees agreed that additional processing of the data such as co-adding WISE and NEOWISE and extracting a multi-epoch photometric database and parallax and proper motion catalog would greatly improve the scientific results of the most important projects identified. In that respect, a project such as MaxWISE which would execute the most important additional processing and extraction as well as make the data and catalogs easily accesssible via a public portal was deemed extremely important.

\end{abstract}

\section{INTRODUCTION}
NASA's Wide-field Infrared Survey Explorer (WISE)\footnote{http://www.jpl.nasa.gov/wise/} was launched into orbit five years ago, and surveyed the entire mid-infrared sky in 2010. The satellite resumed surveying as NEOWISE\footnote{http://neowise.ipac.caltech.edu/} in December 2013, and is now on its fifth pass over the sky. Over 1,200 
refereed papers based on the data have now been published\footnote{http://tinyurl.com/WISEpapers}, with results ranging from the discovery of the first Earth Trojan asteroid, to challenging the standard paradigm for quasar unification.  To celebrate what has been done with WISE, what is being done with NEOWISE, and what will be done in the future, the "WISE at 5: Legacy and Prospects" conference\footnote{http://wise5.ipac.caltech.edu} was held Feb. 10 - 12, 2015, on the campus of Caltech in Pasadena, CA.

\begin{figure*}[!ht]
\begin{center}$
\begin{array}{cc}
\includegraphics[width=3.5in]{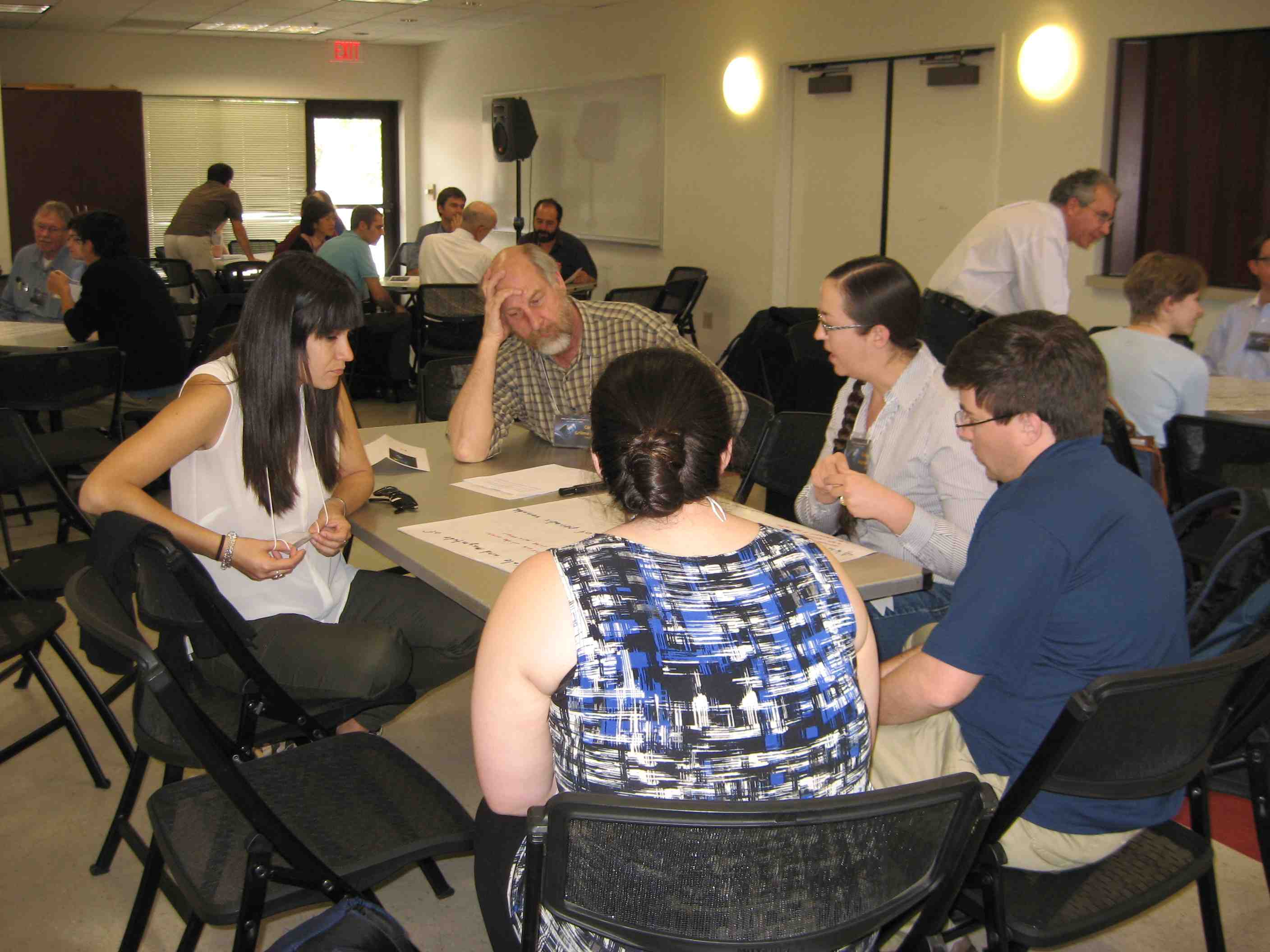}&
\includegraphics[width=3.5in]{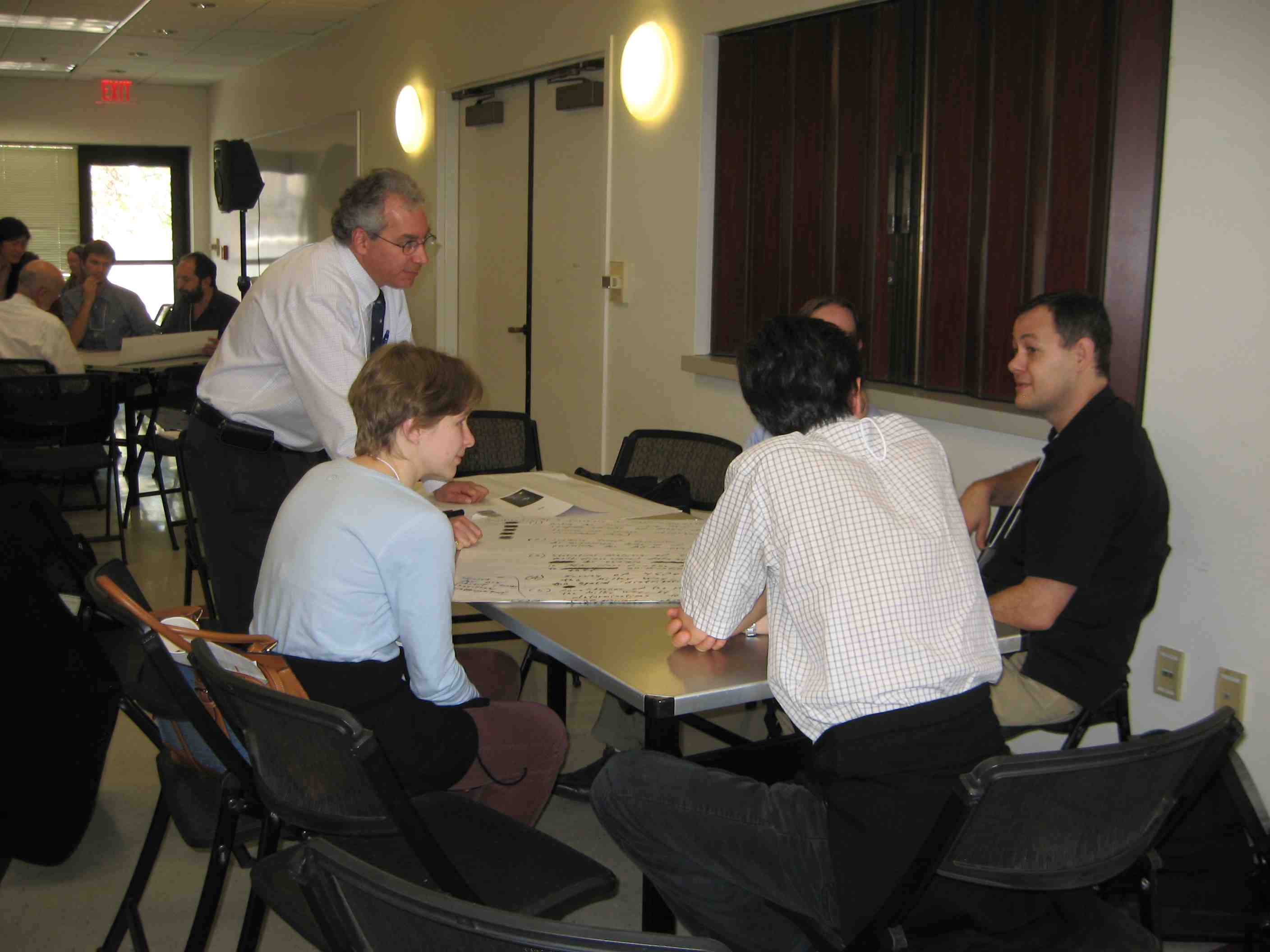} \\
\includegraphics[width=3.5in]{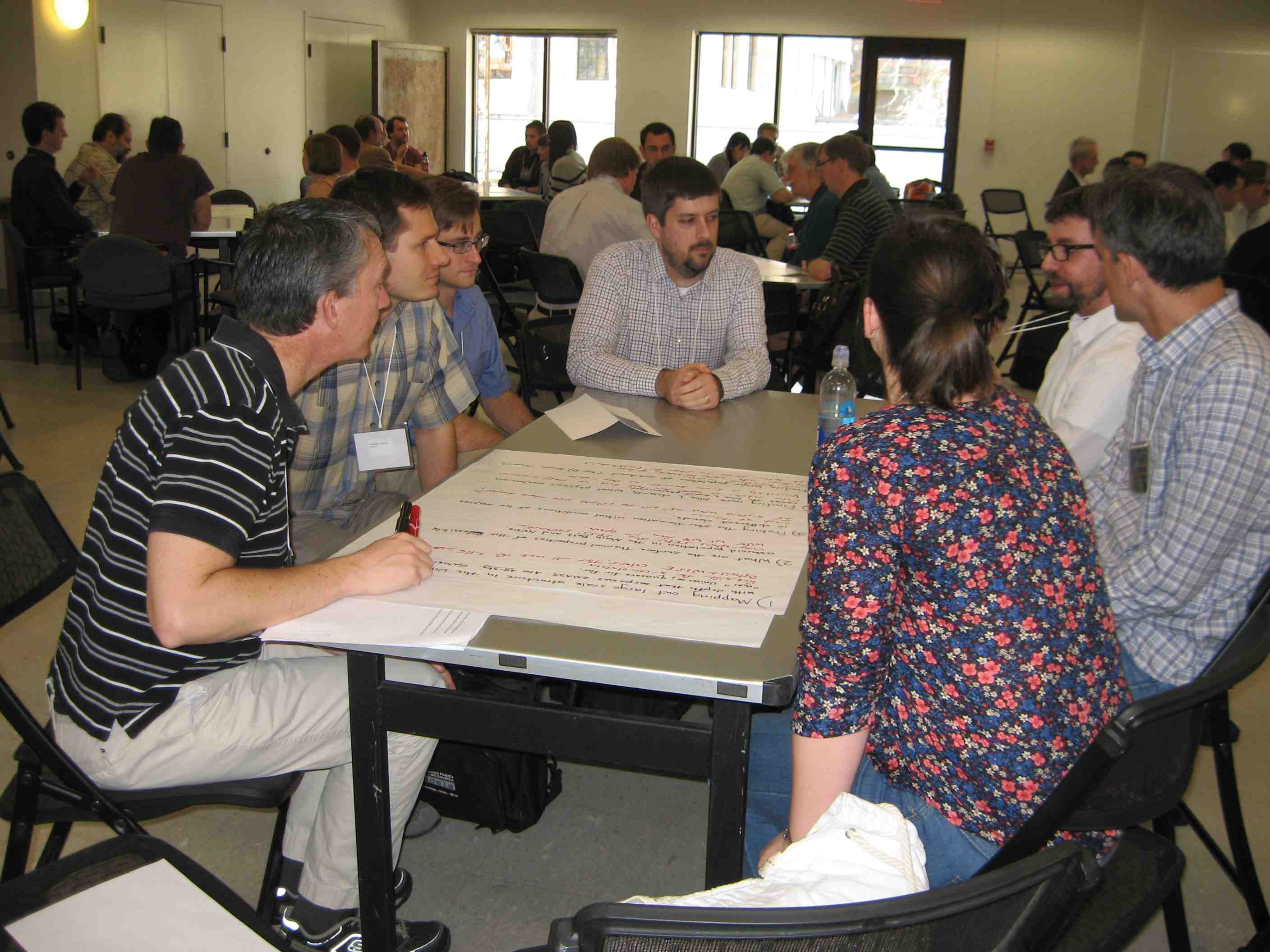} &
\includegraphics[width=3.5in]{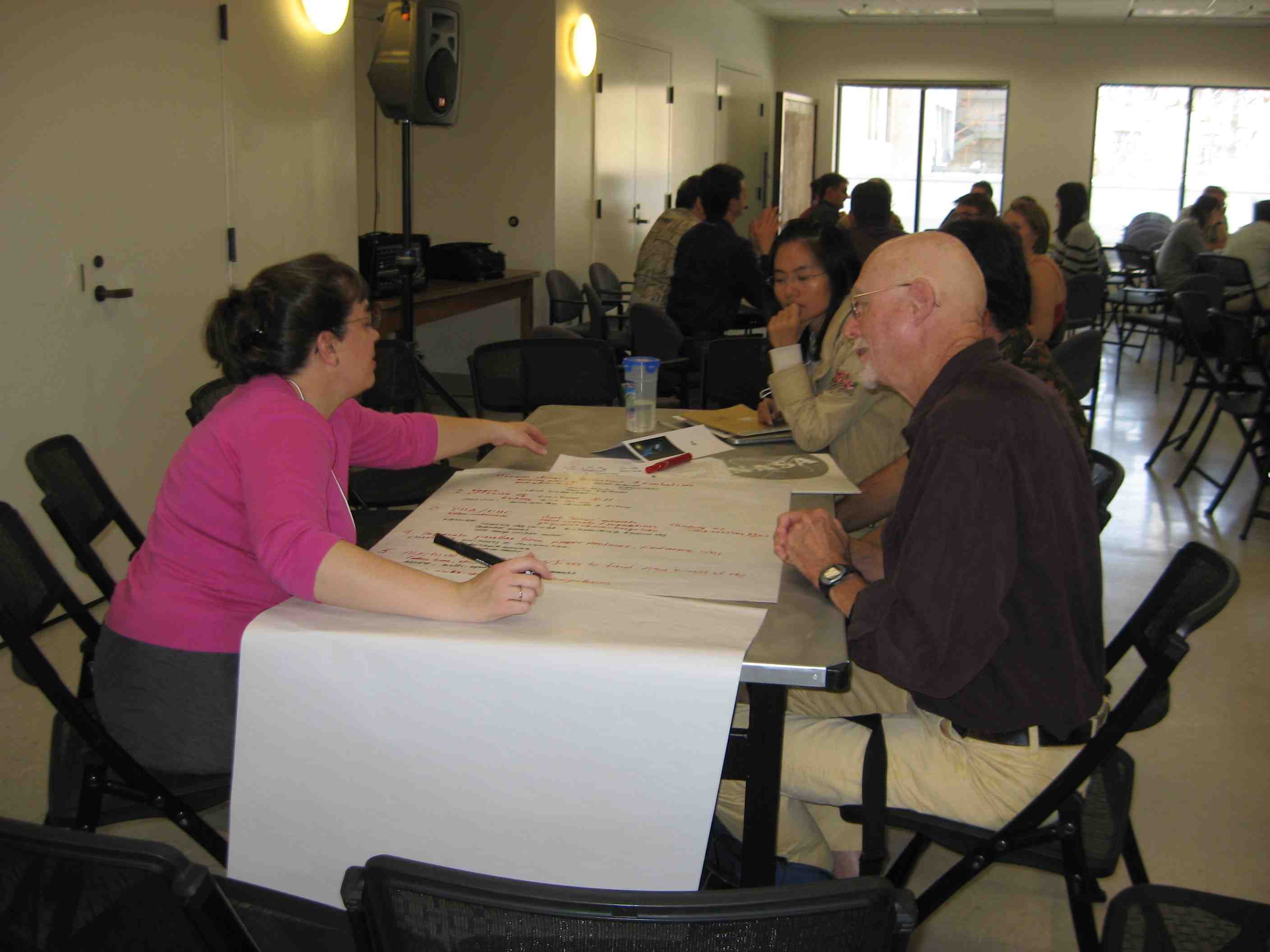} \\
\end{array}$
\end{center}
\caption{Photos taken while the interactive session was in progress demonstrating the organization of the room and the engagement at different tables. }
\label{fig:roomview1}
\end{figure*}
During the meeting, a new type of session was introduced to encourage broad engagment among meeting participants.  In general, scientific meetings offer standard opportunities to interact with fellow attendees.  Typically these happen formally during poster sessions, post-talk question and answer time, or panels. Informally these happen at coffee breaks, lunches, or dinners where attendees are able to isolate key people to engage with individually. However it is often difficult to generate discussions that engage all participants, especially junior attendees who may be intimidated by speaking up during sessions or are shy in informal situations. Recent meetings have been inserting interactive sessions to try and encourage more informal engagement among a larger and broader range of participants (e.g. at the ``Exoplanets and Brown Dwarfs: Mind the Gap'' meeting held in Hertfordshire, UK and at the ``Gaia and the Unseen: the Brown Dwarf Question'' meeting held in Torino, Italy). 
\begin{figure*}[!ht]
\begin{center}$
\begin{array}{cc}
\includegraphics[width=3.5in]{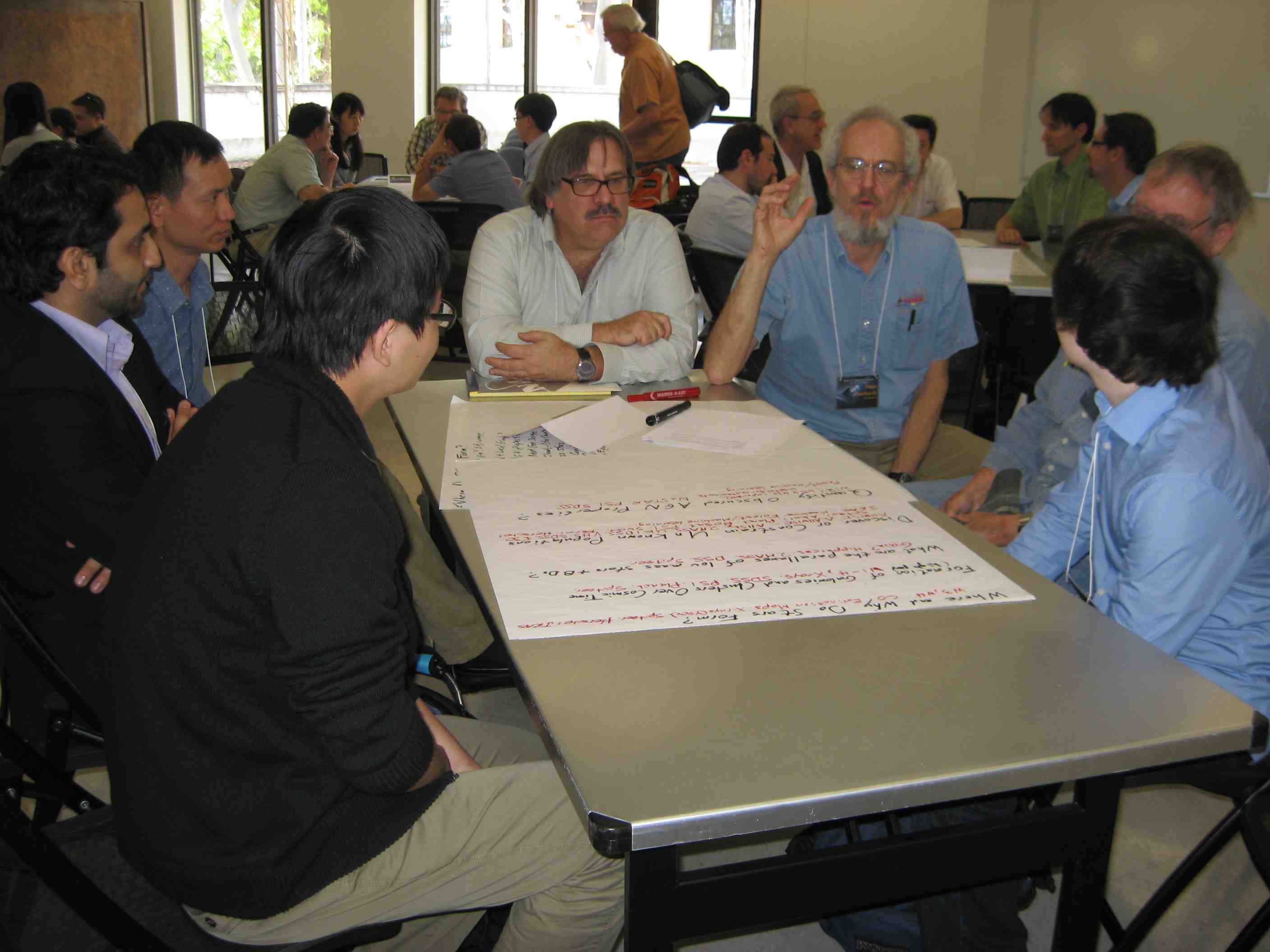}&
\includegraphics[width=3.5in]{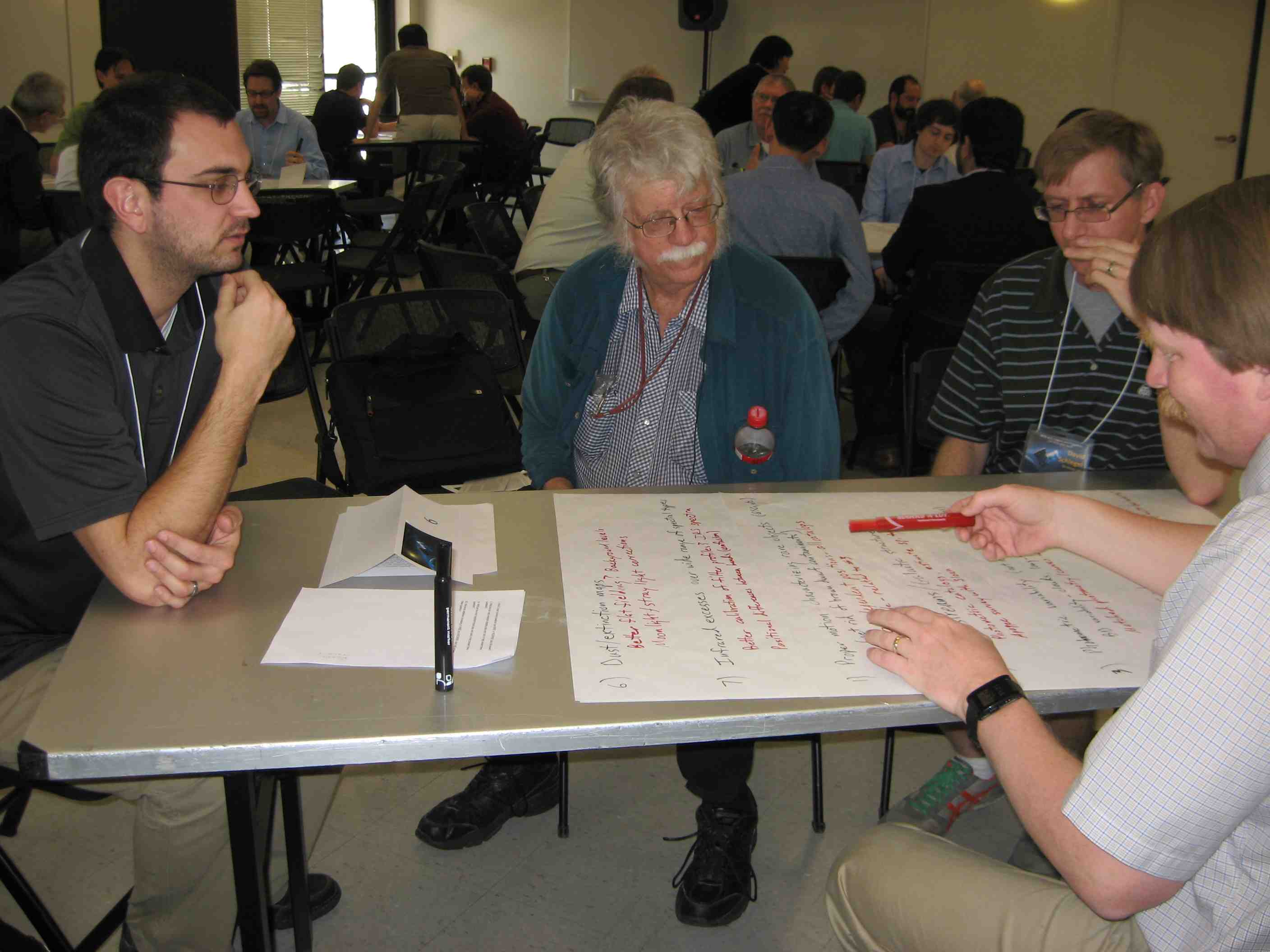} \\
\includegraphics[width=3.5in]{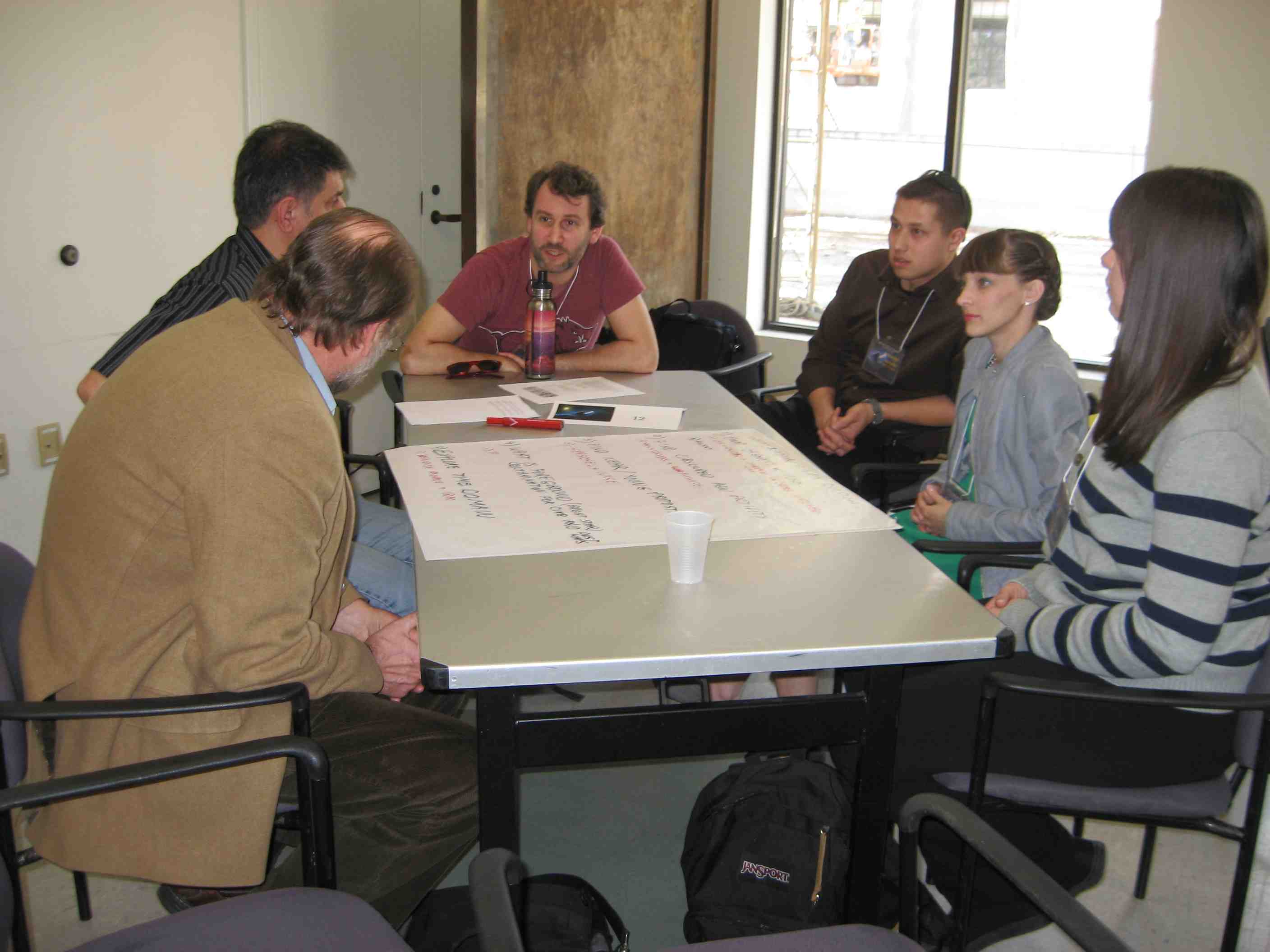} &
\includegraphics[width=3.5in]{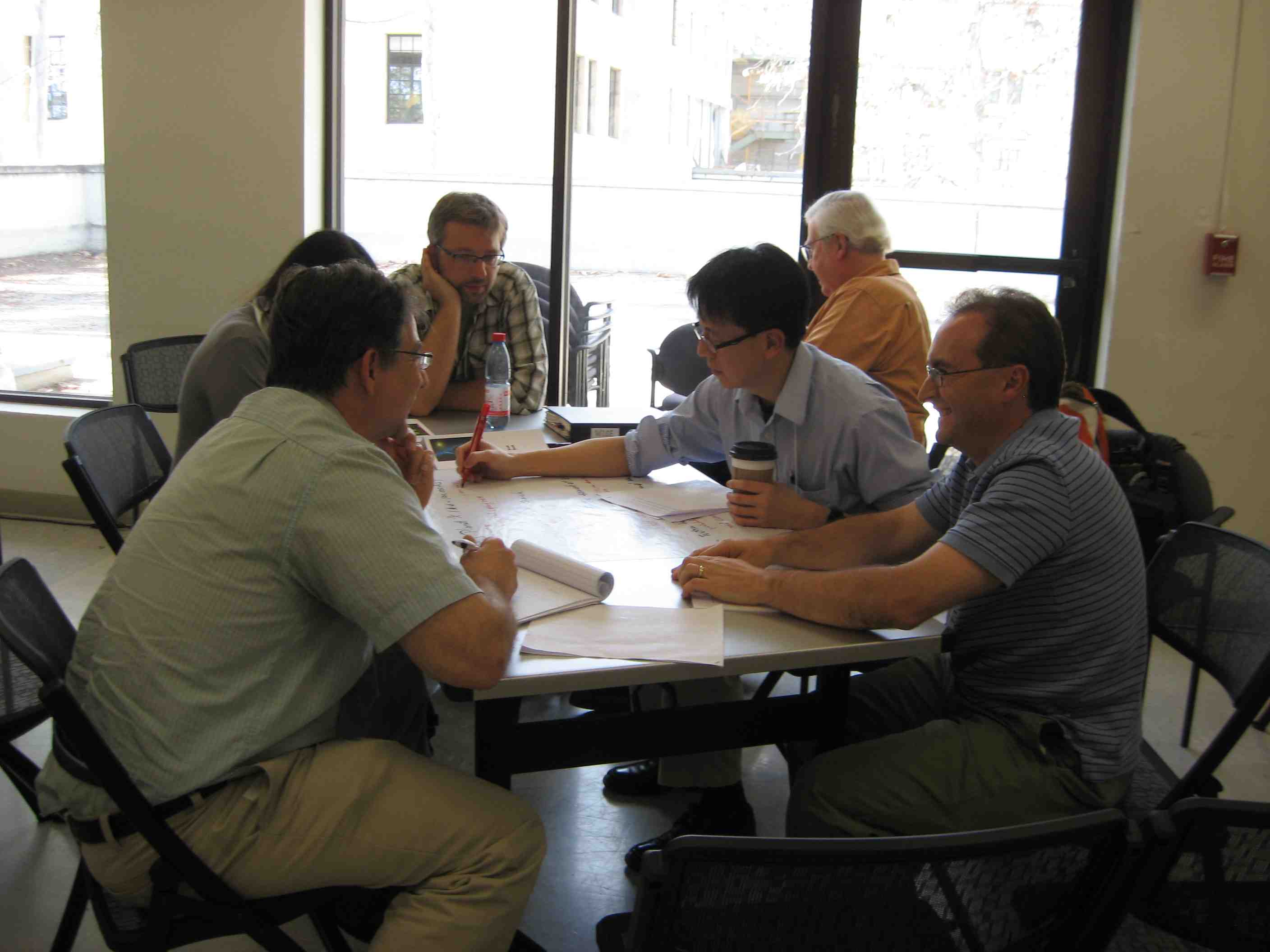} \\
\end{array}$
\end{center}
\caption{Additional photos of the interactive session in progress. }
\label{fig:roomview2}
\end{figure*}
The second day of the ``WISE at 5'' meeting was centered around the topic ``future uses of WISE/ NEOWISE data.'' This type of forward-thinking session, which lacks traditional results or data to present, naturally lends itself to a group discussion. Hence, 90 minutes were dedicated to discussing and brainstorming future uses of WISE/NEOWISE mission data in small, rotating groups. In this paper, we report the results of this interesting session where conference attendees assessed the most important questions that WISE/NEOWISE could still answer as well as the current and future data and tools that would help address these topics. We were particularly interested in assessing what additional processing of WISE/NEOWISE data might contribute to the important scientific questions that the mission would be engaged in during the coming years. In section 2 we describe the procedure followed for setting up and executing the interactive session. In section 3 we summarize the results broken down by each of the four brainstorming questions. In section 4 we present conclusions.

\begin{figure*}[!ht]
\centering
\vskip-0.8in
\includegraphics[scale=0.6]{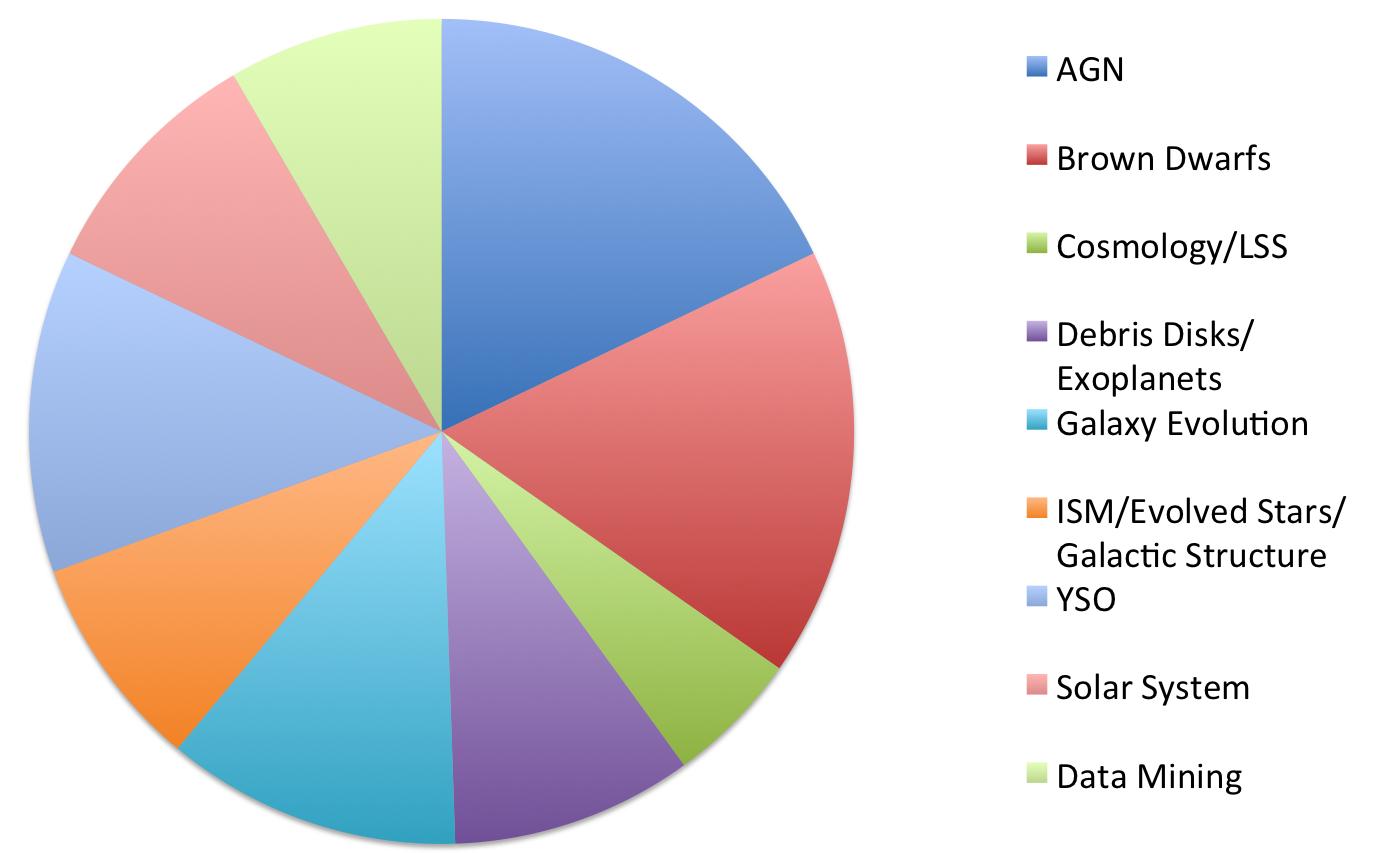}
\caption{The scientific interest distribution of attendees at the ``WISE at 5'' meeting}
\end{figure*}
\section{PROCEDURE}
The plan for the interactive session was to break attendees into small groups and have them brainstorm answers to questions regarding the future uses of the WISE/NEOWISE data (see Figures 1 - 2).   There were $\sim$120 attendees at the meeting from North and South America, Europe, Asia, and Africa.  A large number of North American attendees were local to southern California and therefore the actual number of people present for any given session on any given day varied. Not all conference attendees participated in the interactive session as it was held in a different venue.  In all there were $\sim$65 participants in the interactive session.  

Prior to the meeting, attendees were randomly pre-assigned into 12 groups of up to 10 people\footnote{As organizers were unaware how many attendees would participate in the interactive session, groups ended up ranging in size from 5 - 10 at any table.}. One moderator for each table was contacted ahead of time and asked to remain at a pre-assigned table, help guide the conversation and take notes while other participants rotated.  The session was held in a single, large room with enough tables and chairs to hold all attendees.  Large easel-sized paper and markers were placed at each table so moderators could record group answers to each discussion question.

Before the meeting, the scientific organizing committee developed four questions aimed at engaging attendees on the future uses of WISE and NEOWISE data. Each question built upon the previous. In total, there were 90 minutes dedicated to the session or roughly 20 minutes for each question. Groups began at one table to answer the first question but then rotated in a pre-assigned manner (labeled on their conference badges) to three subsequent new groups. In so doing, attendees engaged with four random groupings of participants. 

\noindent
The questions proceeded in the following order:\\
\noindent
{\bf 1) What five major science questions can WISE or NEOWISE data still answer?}
Participants were given 20 minutes to introduce themselves to their fellow tablemates and brainstorm. At the end of the time, attendees were rotated to a second pre-assigned table and asked to brainstorm the follow-up question: \\
\noindent
{\bf 2) What are the most important existing data and tools to help answer these questions?}\\
\noindent
Participants were at a new table, and therefore were answering question 2 in regards to what a previous group had brainstormed. By having one group work off of what the previous group produced, the conference indirectly engaged at a large level. After another 20 minutes, attendees rotated to a third table with an entirely new pre-assigned group and brainstormed a third question:\\
\noindent
{\bf 3) What are the most important future data and tools to help answer these questions?}\\
\noindent
After another twenty minutes of brainstorming on the answers provided by group 1 and 2 at a given table, the attendees rotated one final time to answer a fourth question:\\
\noindent
{\bf 4)\footnote{We acknowledge that Q3 and Q4 are somewhat degenerate, but we considered the question about upcoming WISE/NEOWISE data products to be important enough to require a separate answer.} How would additional processing of WISE and NEOWISE data help answer these questions?}\\
\noindent
In this manner the output of any one table was the product of four rotations or dozens of attendees brainstorming. 
\begin{figure*}[!ht]
\centering
\includegraphics[scale=0.4]{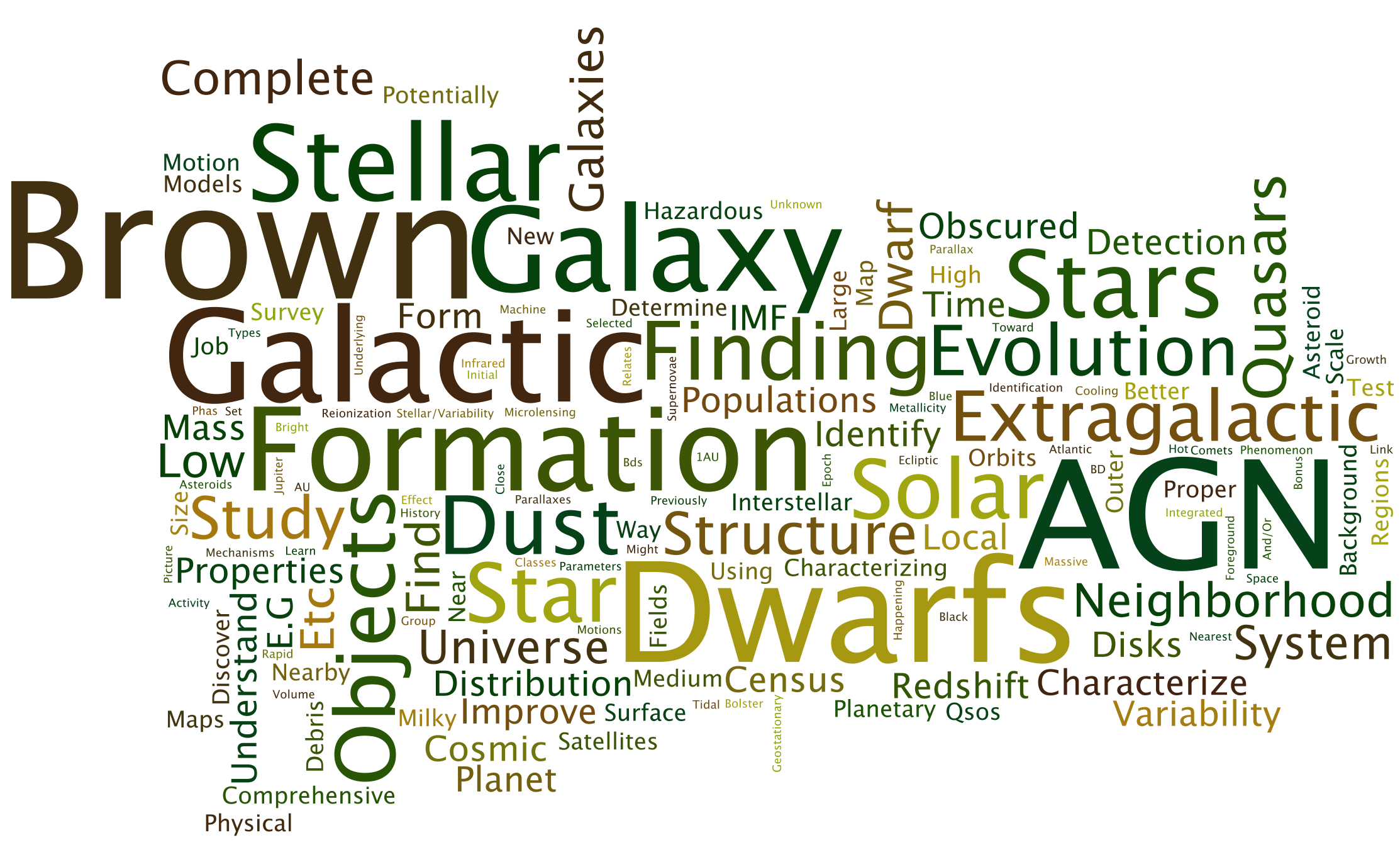}
\caption{A word cloud of the moderators' summaries of the 64 scientific questions that WISE or NEOWISE will tackle in the coming years as brainstormed by conference participants. Word size is roughly proportional to the frequency of mentions. }
\end{figure*}
Moderators were anchors at each table and did not rotate but rather helped guide discussions. To ensure that we were able to collate the results from the attendees, we asked the moderators to take careful notes during the session. Furthermore, since the point was to rotate participants through different tables, it was important to ensure that the work of a previous rotation was clearly visible. Hence each table was given a giant easel-sized paper and the responses to each question were boldly displayed. Post session, moderators were asked to summarize the results of their table using an online document. We then collated the responses and asked the moderators to comment in more detail (as described in the summary below).

\section{SUMMARY}
Many attendees of the ``WISE at 5'' conference were extensive users of the dataset and the results reported here will reflect the scientific interests of the participants. However, the expertise of attendees spanned a range of topics -- from solar system objects to quasars (see Figure 3 for a distribution of the scientific interests of the attendees). Therefore, individual bias should be minimized\footnote{We note that the NASA Discovery proposal deadline coincided with the meeting therefore a number of planetary scientists were unable to participate in the session.  Consequently there is a lack of emphasis on projects related to this important scientific focus of the WISE/NEOWISE mission.}. Participants in the interactive session were asked to think both within and beyond their own field and they were encouraged to think outside the box.

\subsection{What five major science questions can WISE or NEOWISE data still answer?} 
The first brainstorming session was the set-up for all subsequent interactions. Participants were asked to work with their first assigned table and come up with five major science questions that WISE or NEOWISE could still answer. We allowed 20 minutes for group discussions. The outcome was 64 WISE-relevant science questions that were submitted across 12 tables (two tables submitted six questions and one table submitted seven). See Figure 4 for a word cloud of all the responses. These questions spanned a range of scientific topics and reflected the expertise and forward thinking of the attendees. The questions brainstormed ranged from broad (e.g. Where and why do stars form? and How do galaxies and clusters form and evolve over cosmic time?) to specific (e.g.  Study the integrated Sachs-Wolfe effect in the local Universe, and Study the structure of magnetic fields in the interstellar medium). We also asked the attendees to consider ``outside the box'' ideas and several responses fell into that category (e.g. Characterize the South Atlantic Anomaly, and Characterize geostationary satellites). 

\begin{figure*}[!ht]
\centering
\includegraphics[scale=0.5]{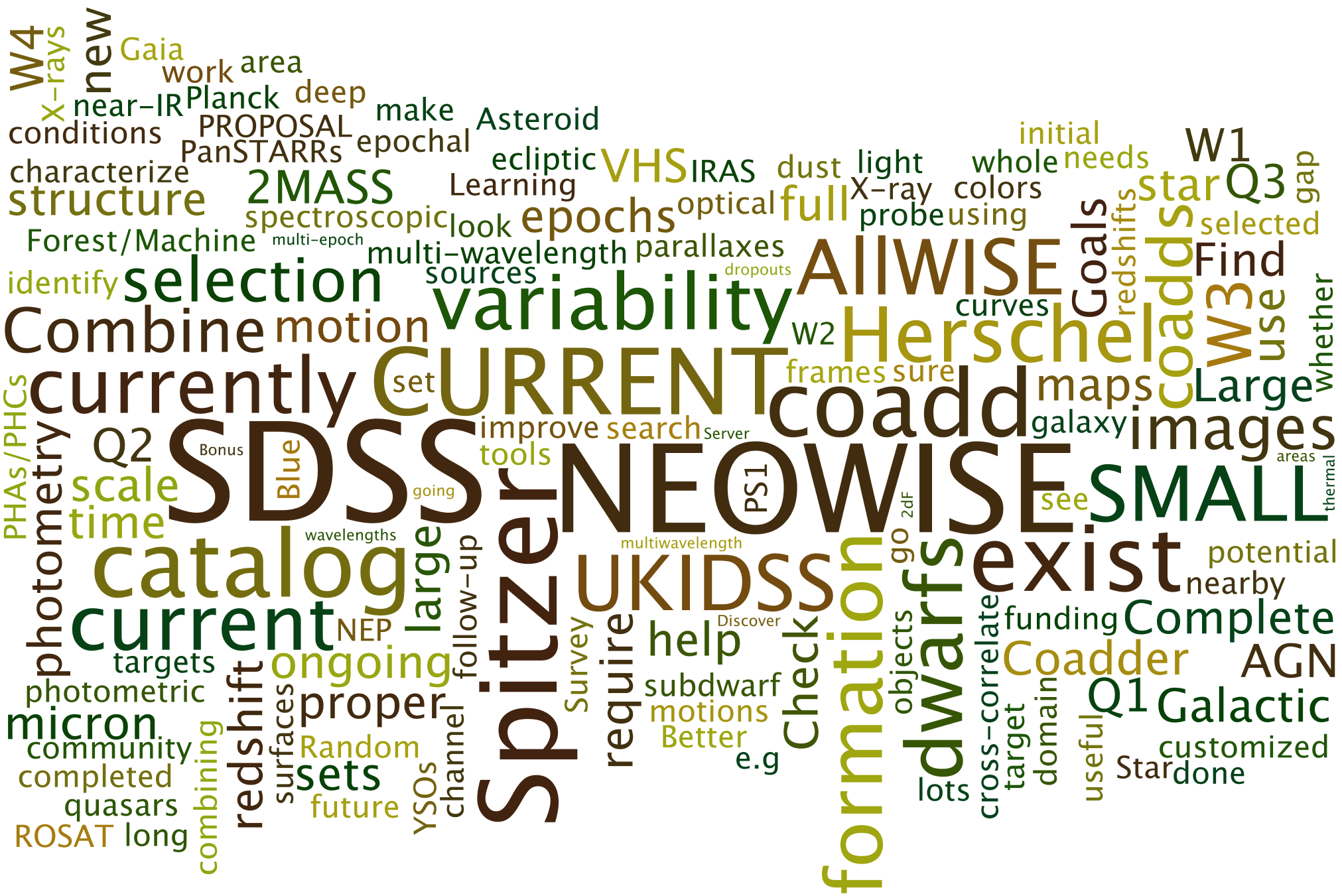}
\caption{A word cloud of the moderators' summaries of the current tools that will help answer the scientific questions brainstormed in session \#1. }
\end{figure*}
We found the majority of answers were bifurcated between two themes: (1) brown dwarf questions, and (2) active galactic nuclei (AGN) or quasar questions. There were six WISE-relevant science questions repeated more than once across the 12 tables. We list each of these along with the number of times that they appeared in parentheses below:\\
\vskip-0.2in

\noindent
(1) Determine the initial mass function (IMF) in the substellar regime (x6)\\
(2) Identify obscured AGN and test AGN unification (x5)\\
(3) Discover new quasars (x3)\\
(4) Discover or constrain unknown populations (x3)\\
(5) Characterize high redshift quasars (x2)\\
(6) Find Planet X (x2)\\

\vskip-0.2in
To summarize the 43 remaining results, we grouped the questions by common scientific theme. The following were recurring topics that appeared at least two times: \\
\vskip-0.2in
\noindent
(1) Exploring populations of brown dwarfs (x6)\\
(2) Galactic structure (x5)\\
(3) Star formation (x4)\\
(4) Large scale structure (x3)\\
(5) Zodiacal dust (x2)\\
(6) Asteroid populations (x2)\\
(7) Man-made satellites (x2)\\
(8) Dust extinction maps and the cosmic microwave background (x2)\\
\vskip -0.1in
\noindent
For those interested, we have made the full list of WISE-relevant science questions available to view in a shared Google document\footnote{http://tinyurl.com/o74vzpy}. 

\subsection{What are the most important existing data and tools to help answer these questions?} 

The second brainstorming session was focused on evaluating what {\bf current} instrumentation, facilities and/or tools would help address the scientific questions enumerated in the first session (see Figure 5 for a word cloud of the discussions). In the moderators' summary of each table's response, we asked them to generalize if (1) the follow-up data already existed (e.g. all-sky photometric surveys) or (2) if one needed to apply for the time. In the case of the latter, we also evaluated whether or not this follow-up could be obtained by small proposals (i.e. individuals could spearhead the work) or if it required a large collaborative follow-up (i.e. something the community would need to ask for). 

In general, many of the scientific questions cited cross-correlating WISE data with other all-sky or large-area surveys such as SDSS\footnote{The number of questions for which each survey was cited is listed in parentheses.} (x7), 2MASS (x4), UKIDDS (x4), and VHS (x3) as well as those conducted by $Spitzer$ and $Herschel$ as significant for follow-up. In that respect many groups converged on the idea that multi-wavelength photometric approaches (using the above surveys as well as ROSAT--x3, Fermi-x2, Swift-x2, GALEX--x2) would help with the most important science questions. Co-adding the current WISE data and extracting parallaxes, time-variations, and proper motions was also cited although this is discussed in greater detail in section 3.4. 

The major follow-up using current data that requires a community effort to execute was in regard to spectroscopic follow-up. This was noted for several of the scientific goals that varied in theme including: (1) investigating the initial mass function into the substellar regime, (2) exploring AGN unification, and (3) establishing the large-scale structure map of the Universe over the entire sky. 

A few current tools of note that would aid in addressing the scientific topics were (1) the IRSA finder chart tool, (2) the WISE image server, and (3) the SQL interface for AllWISE. Also noted as critical for achieving many of the scientific goals was an implementation of machine-learning algorithms. This was of note for both current and future uses as teams were not sure whether these were tools that needed to be developed or if they already existed.

\subsection{What are the most important future data and tools to help answer these questions?}
\begin{figure*}[!ht]
\centering
\includegraphics[scale=0.5]{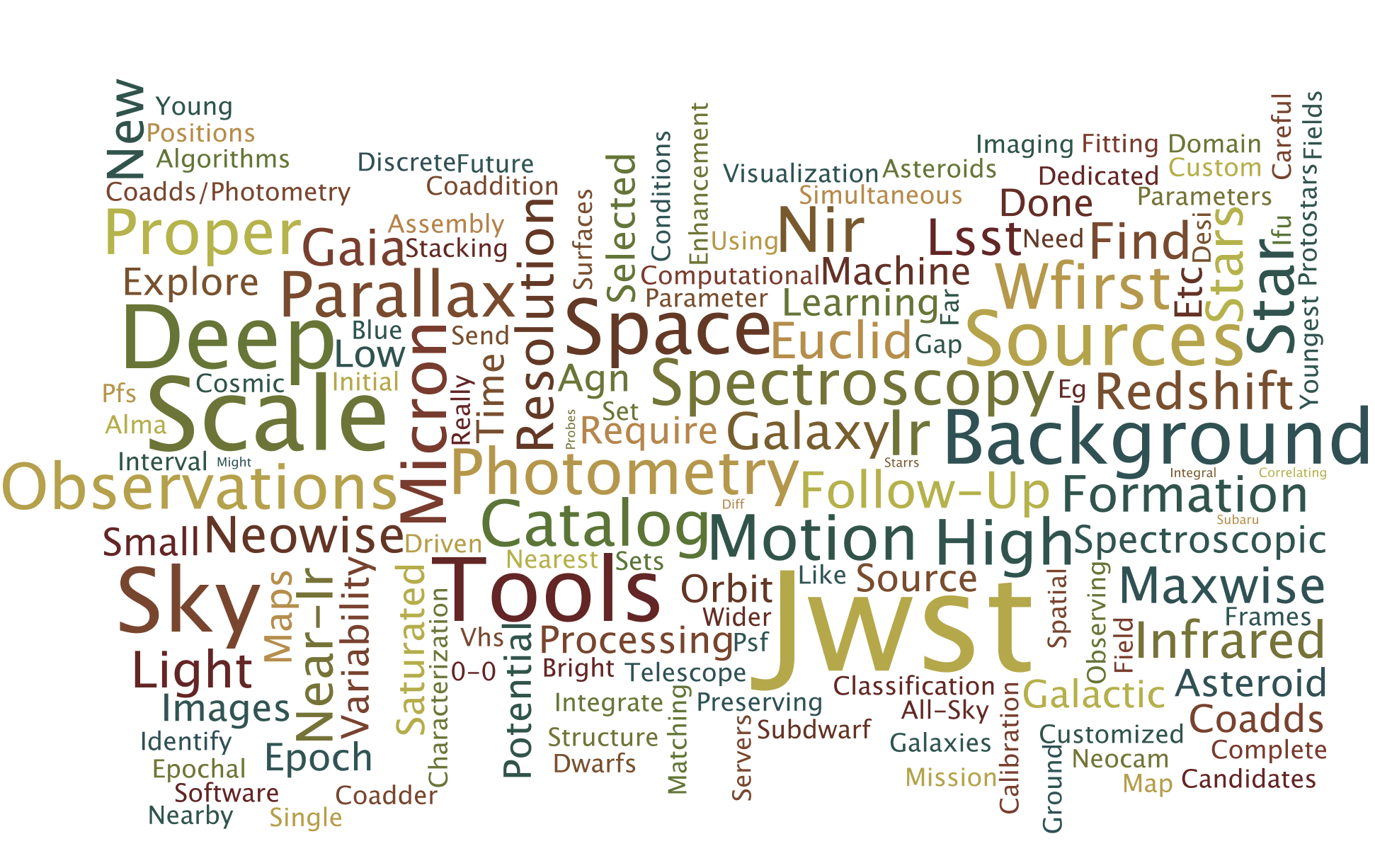}
\caption{A word cloud of the moderators' summaries of the most important future data and tools that will help answer the questions brainstormed in session \#1. }
\end{figure*}

The third brainstorming session was focused on evaluating what {\bf future} instrumentation, facilities and/or tools would help address the scientific questions enumerated in the first session (see Figure 6 for a word cloud of the discussions). As with brainstorming session \#2, we asked the moderators to summarize by reflecting on whether or not the future follow-up could be done with small proposals (i.e. individuals could spearhead the work) or if it would require a large collaborative follow-up (i.e. something the community would need to ask for). 

Several upcoming facilities were overwhelmingly present among the desired follow-up for addressing the scientific questions. Those included spectroscopic follow-up using the James Webb Space Telescope (JWST) as well as the Wide-Field InfraRed Survey Telescope (WFIRST-AFTA) and a cross-correlation with the data from the Large Synoptic Survey Telescope (LSST) as well as Gaia and Euclid. As was described in the session \#2 summary, many groups stated that complementing the WISE data with multiwavelength photometric surveys (including VHS, and TESS in addition to those listed above) would be beneficial for the most important scientific questions. 

In regards to whether small proposals or a community effort would be required, there was a similar response as to that detailed in the summary of session \#2. Small proposals on major future instrumentation were highly regarded although several projects, such as an all-sky spectroscopic survey to determine high redshift sources, were also cited. 

As multiwavelength astronomy and cross-catalog searching was very high on the list of responses, several groups noted that tools to facilitate large-scale catalog matching across a broad list of surveys would augment the science results moving forward. Moreover, developing and implementing machine-learning tools to explore and understand the data were often mentioned.

\subsection{How would additional processing of WISE and NEOWISE data help answer these questions?}
For the fourth and final brainstorming session, we asked participants to comment on how additional processing of WISE and NEOWISE data would help answer the questions enumerated in session \#1. Across brainstorming sessions \#2 and \#3, we found that the co-adding of WISE/NEOWISE data as well as source extractions for motion, parallax, and variability searches were cited multiple times as both a current and future tool for addressing the science questions. Hence, without even asking this follow-up question, we had an overwhelming consensus that additional processing of WISE and NEOWISE data would aid in answering the most important questions possible with the mission.

In the moderators' summary of this session, we asked for (1) a description of how the additional processing would impact the scientific project as well as (2) specifics on the type of processing discussed at each table. Furthermore, moderators were asked to report whether the additional processing would require some large funding source (e.g. MaxWISE\footnote{See poster by P. Eisenhardt from the ``WISE at 5'' meeting\\ http://wise5.ipac.caltech.edu/posters/Eisenhardt.pdf
}) or whether individual ADAP grants could cover it. 

To summarize, the additional processing most often recommended was co-adding the WISE and NEOWISE data to facilitate the discovery of fainter objects (e.g. QSOs, AGN, Y-type brown dwarfs, galaxy clusters). Moreover, the creation of a complete and detailed Multi-epoch Photometry Database (MEP), as well as a proper motion and parallax catalog, dominated the needs of many science programs (e.g. identifying unknown populations of brown dwarfs, finding RR Lyrae variables to determine the distance to and structure of tidal streams). Also of note were (1) better background subtraction which would help with both Galactic and extragalactic projects (e.g. the detection of low surface brightness light from ejected stars, detection of diffuse emission in stars and dust to map the Milky Way in 3D), and (2) an improved PSF model which would help with the numerous astrometric and variability studies. 

As to the question of whether this additional processing would require a large funding source or whether individual ADAP grants could cover it, there was a diversity in responses. MaxWISE, the proposed co-adding of WISE data and catalog extraction, was cited as important for numerous questions. This type of activity would enable the main additional processing points that were cited by attendees as critical for addressing the most important science questions possible with the WISE mission. In the summaries, the moderators noted that portions of the additional processing, such as co-adding parts of the sky of interest or extracting only certain regions, were certainly plausible with something like an ADAP grant. However, what seems clear from this session is that a concerted effort to uniformly co-add the data and extract sources is the optimal product for addressing the most important science questions that WISE/NEOWISE can still answer.  Importantly, many noted that a project like MaxWISE was required to both execute the additional processing and extraction AND make the end result accesible to the community.

\section{CONCLUSIONS}
The ``WISE at 5'' meeting brought together experts and super-users of the dataset. As such, it was an excellent opportunity to identify the major scientific questions that could be answered by WISE/NEOWISE in the coming years as well as the complementary datasets, tools, or additional processing that would aid in addressing these topics. Overall, the results showed several recurring topics bifurcated between extragalactic and brown dwarf science. Likely this is a reflection of the scientific preference of the attendees. 

Among the 12 separate tables asked to brainstorm five important (or potentially out of the box) questions that WISE could answer in the coming years, several questions came up repeatedly, including: (1) Determine the initial mass function (IMF) in the substellar regime; (2) Identify obscured AGN and test unification; (3) Characterize high-redshift quasars; and (4) Find Planet X. Many of the scientific questions will be complemented by a cross-correlation with other photometric surveys (e.g. SDSS, Pan-STARRS, LSST, Gaia, Euclid, etc.) and would benefit from the development or implementation of machine-learning algorithms to aid in the studies. 

Finally, it was largely agreed by attendees that additional processing of WISE data would greatly benefit all future studies. Co-adding WISE with NEOWISE data, extracting a multi-epoch photometric database as well as a proper motion and parallax catalog that would be easily accessible in a public portal was among the top cited additional processing techniques. While small grants (e.g. ADAP) would enable some of the work, a community-backed project such as MaxWISE was mentioned numerous times and would greatly improve our ability to address the most important scientific topics cited by conference participants.

\end{document}